\documentclass[reprint,nofootinbib,amsmath,amssymb,aps,]{revtex4-1}

\usepackage{graphicx}
\usepackage{hyperref}
\usepackage{xcolor}

\begin{document}

\title{You Have to Grow Wefts to Fold Them}

\author{Lauren Niu}
\affiliation{Center for Functional Fabrics, Drexel University, Philadelphia, PA}
\affiliation{Department of Physics and Astronomy, University of Pennsylvania, Philadelphia, PA}
\author{Genevi\`eve Dion}
\affiliation{Center for Functional Fabrics, Drexel University, Philadelphia, PA}
\author{Randall D. Kamien}
\affiliation{Department of Physics and Astronomy, University of Pennsylvania, Philadelphia, PA}

\date{\today}

\begin{abstract}
Knitting can turn a one dimensional yarn into a highly ramified three-dimensional structure.  As a method of additive manufacturing, it holds promise for a new class of lightweight, ultrastrong materials.  Here we present a purely geometric model to predict the three-dimensional self-folding of knitted fabrics made only of the two traditional stitches, knit and purl.
\end{abstract}

\maketitle

\section*{Introduction}

It is undeniable that, by any metric, geometry plays a central role in our understanding of the physical world.  From celestial mechanics \cite{newton1687philosophiae} to soap bubbles \cite{thomson1887lxiii} and from optics \cite{maxwell1867on} to gravity \cite{misner1973gravitation}, the rigor of geometric logic intoxicates our thinking.  Indeed, because of their intrinsic elegance and potential application, tools have been developed to make materials that assemble into a targeted topography.  Whether it be through the techniques of \textit{origami} \cite{lang2012origami, yoshizawa2016akira, demaine2017origamizer}, thermal activation of local material anisotropy \cite{klein2007shaping}, or pneumatic actuation \cite{siefert2020programming}, the common thread is the use of isometric or near-isometric embeddings.  Knit materials, on the other hand, have relatively small elastic moduli and can be designed by their creator to fold into complex three-dimensional patterns upon their construction.  Here we propose a purely geometric model that can rationalize the folding of a myriad of knit motifs, and provide direct qualitative comparison between our simulated results and knitted fabrics.

\subsection*{Self-Folding Knits}

\begin{figure}[h!]
    \centering
    \includegraphics[width=0.45\textwidth]{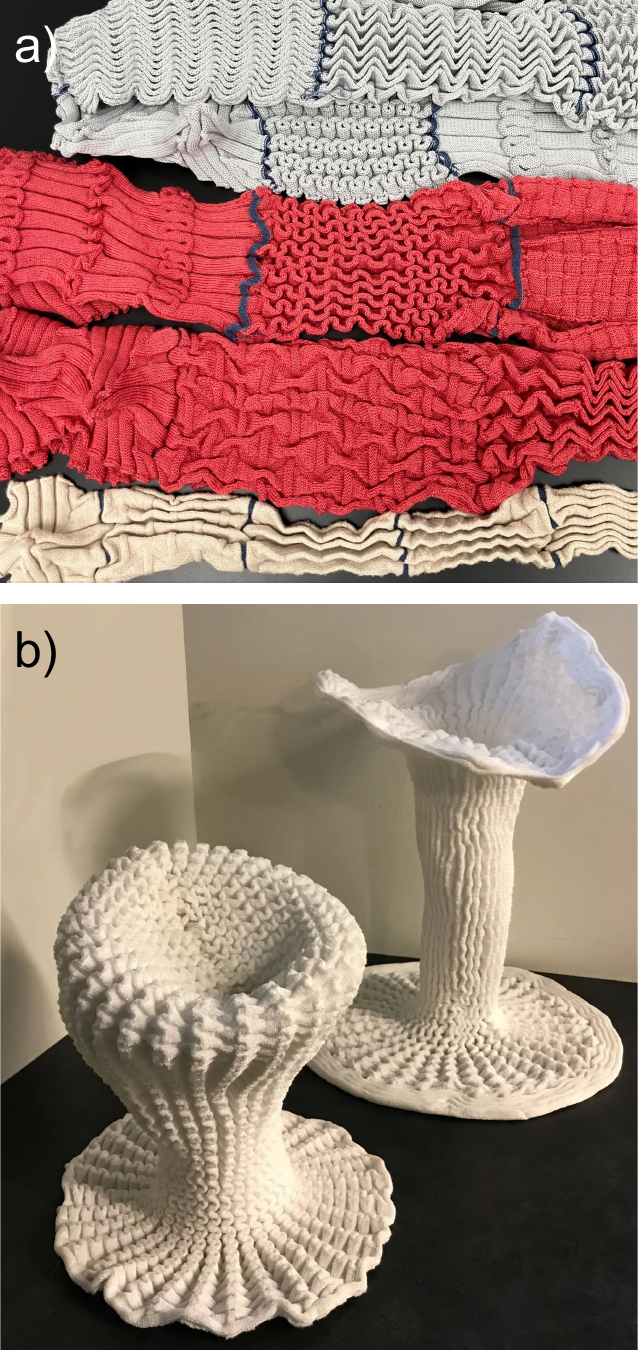}
    \caption{
        a) Samples of periodic self-folding weft knit patterns, knitted by machine from three different types of yarn. From top to bottom: recycled REPREVE® nylon yarn, Jaguar Modal/Nylon yarn, and Ecocot cotton yarn.
        b) Weft knits with stitch patterns that have non-flat topology (left: elliptical, right: cylindrical) can also be used to shape large 3D forms. Samples were knitted with a nylon/PET blended yarn and steamed after fabrication to rigidify the fabric. Figure reproduced with permission from \cite{dion2024knitogami}.
    }
    \label{fig:self_folding}
\end{figure}

Weft-knit materials showcase a wide variety of complex geometric and mechanical behavior \cite{rant2013foldable, pavko2017multifunctional, poincloux2018geometry, amanatides2022characterizing, singal2024programming}. The 3D structure of these patterns typically requires no or minimal post-processing after initial fabrication, and their forms are robust to extensive handling in a variety of environments. Although the exact 3D form generated from a knit pattern may differ based on the choice of yarn and other fabrication parameters, the qualitative form does not depend on these parameters; knitted swatches created at a wide variety of length scales and moduli form similar buckling patterns. Although the focus of our study is on planar weft-knitted fabrics as shown in Figure~\ref{fig:self_folding})a, changing the topology of the whole fabric can yield a variety of different structures with large-scale 3D properties (Figure~\ref{fig:self_folding})b).

The basic stitch of a weft-knitted fabric is shown in Figure~\ref{fig:knit_schematic}a, and is formed by drawing a loop of yarn through a previously existing loop. When the drawn yarn is pulled from ``back to front'' through the previous loop, the resulting stitch is called a \textit{knit} stitch; a loop of yarn drawn through from ``front to back'' creates a \textit{purl} stitch. The structure of knit and purl stitches differ only by a $180^\circ$ rotation, which flips the orientation of crossings in a diagram of the yarn path. The back side of a fabric created with knit stitches is therefore functionally identical to the front side of a fabric created with purl stitches.

This local topological structure of knit and purl stitches, dictated at fabrication, creates the overall tendency for a knitted swatch to curl dramatically (Figure~\ref{fig:knit_schematic}a) \cite{kurbak2008basic, kaldor2008simulating}. This natural curvature is primarily a bulk phenomenon, as the free boundaries of a knit swatch develop large curvatures regardless of the their positions (Figure~\ref{fig:knit_schematic}bc). Knit swatches tend to curl backward in the horizontal direction and forward in the vertical direction; purl swatches do exactly the opposite (Figure~\ref{fig:knit_schematic}a).

\subsection*{From Yarn Topology to Fabric Geometry}

The natural curvatures of a knit or purl fabric can be estimated from the orientation of its yarn crossings in the plane. At any given crossing, the two roughly perpendicular yarns must curve over or under each other, requiring them to bend out of the fabric plane (Figure~\ref{fig:knit_schematic}d). For any yarn with nonzero flexural modulus, this creates a bending moment in the opposite direction. This results in nonzero bending moments at the yarn crossing in the directions of both the top yarn and the bottom yarn, with opposite signs for each. In a region of knit (purl) stitches, all of the crossings align in a similar direction such that the top (bottom) yarn is roughly vertical and the bottom (top) yarn is roughly horizontal (Figure~\ref{fig:knit_schematic}a). This results in a field of aligned bending moments, which drives the observed curvatures and buckling behavior of the fabric.

Most weft-knitted fabrics, whether produced by machine or by hand, are created while keeping consistent tension on the yarn, and typically with the goal of maintaining an even stitch size. We therefore assume that the length of each knitted loop remains roughly the same across the entire fabric, and therefore that the ``area'' of a knitted stitch does not vary significantly over a swatch of 100 or more stitches. Additionally, yarn loops in the final fabric tend to abut one another without leaving gaps in the fabric larger than the yarn width, we observe little distortion in the lengths of knit or purl stitches within a larger swatch post-fabrication, requires overcoming a significant amount of yarn-to-yarn friction \cite{jeddi2004relations, bai2018experimental, poincloux2018crackling}.

The even stitch size, coupled with the condition that we only study flat knit-purl patterns with no topological pattern defects (such as short rows or increases/decreases between rows \cite{popescu2018automated, narayanan2018automatic}), suggests that our final fabrics should have a near-flat metric, even in their self-folded state. We can then invoke Gauss's \textit{Theorema Egregium}, which states that, for an initially flat surface, deformations in three dimensions with nonzero Gaussian curvature (simultaneous curvature in multiple directions, resembling a saddle surface or a spherical dome or bowl) require some regions of the surface to stretch or compress tangentially. Because stretching or compressing a knitted fabric tangentially requires changing the area of stitch loops, we expect that the equilibrium configurations of a knitted swatch will naturally minimize regions of large Gaussian curvature. The result is that rectangular swatches of knit or purl stitches tend to curl cylindrically, which requires no Gaussian curvature, or remain nearly flat (Figure~\ref{fig:knit_schematic}a). Gaussian curvature may still appear in the final surface, albeit at an energetic cost.  

\subsection*{Two Dimensional Surface Approximation}

Our observations of the physical samples suggest that we can approximate the buckled shape of an arbitrary rectangular knit and purl patterns as an elastic sheet with nonzero bending and stretching moduli, imbued with a field of natural curvatures matching the knit/purl pattern. For simplicity, we use an elastic model implementing a F\"oppl-von-K\'arm\'an energy for thin shells with 2D stretching and bending contributions. As minimizing this energy requires solving a global optimization problem that is analytically intractable for our patterns, we use a finite-difference simulation to compare our results with machine-knitted samples.

Because the structure of a knitted swatch differs significantly from the structure of an isotropic elastic sheet, we do not expect the details of the elastic formulation to significantly affect the results. In particular, we do not expect the elastic thickness parameter $h$ used in our simulations to coincide with the physical thickness $t$ of a knitted fabric, although we expect them to be at similar order. Rather, $h$ remains a free tuning parameter that relates the effective bending stiffness of the fabric to its in-plane tensile stiffness. We also note that the physical bending stiffness of a physical knitted fabric may vary based on the stitch patterning and fabrication parameters.

As the elastic sheet approximation by definition ignores the structure of features below the length scale of $h$, which in our case is comparable to the dimensions of a single stitch, we ignore corrections to the elastic sheet energy beyond second order in the in-plane strain or out-of-plane curvature.

\begin{figure*}[h!]
    \centering
    \includegraphics[width=.95\textwidth]{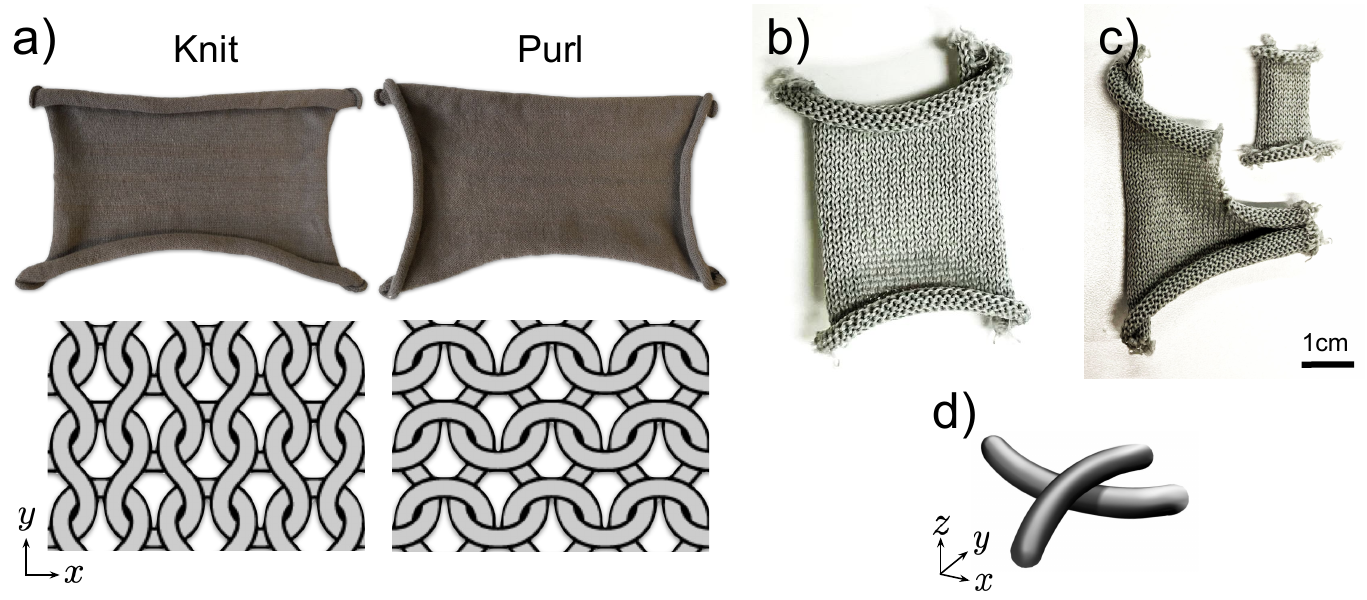}
    \caption{
        a) Two merino wool swatches made entirely from knit stitches and purl stitches (top), and their yarn crossing diagrams (bottom). These swatches exhibit curvature in multiple directions; the knit swatch curls forward at the top and bottom edges, and backward at the sides, while the purl swatch curls in the opposite directions.
        b) A small rectangular piece cut from a larger knit swatch (REPREVE® nylon yarn, approximately $0.5 \text{mm}$ diameter) displays the same rolling behavior as in (a).
        c) When a corner of the knit swatch in (b) is cut away, all free boundaries continue to curl significantly. This implies that the edge curling behavior observed in knit swatches is likely due to a \textit{bulk} rather than a boundary stress.
        d) At any yarn crossing in a textile, the two strands must bend around each other to avoid physically intersecting. This results in a local bending moment that is positive in the $y$-direction and negative in the $x$-direction. The field of bending moments generated from each yarn crossing contributes to the overall geometry of the fabric.
    }
    \label{fig:knit_schematic}
\end{figure*}

From experimental observations in Figure~\ref{fig:knit_schematic}a-c, small swatches of knit materials with dimensions of approximately 40 stitches per side length show large natural curvatures both in the vertical and horizontal directions. We do not measure these natural curvatures physically, but we note that all free boundaries of fully-knit or fully-purl samples naturally rolled up tightly enough that their curvatures were limited by self-intersection of the fabric, such that $|\kappa|_x \sim |\kappa|_y \sim 1/t$, where $t$ is the thickness of the fabric.

\section*{Results}

Figure~\ref{fig:knit_patterns} shows a comparison of several simulated fabrics with experimental swatches knitted from nylon REPREVE® yarn using the same pattern. We use a quasistatic method for simulation to grow the patterns, starting from a flat sheet and incrementally increasing the magnitudes of $\kappa_x$ and $\kappa_y$ at every step, using each step's solution as the initial condition for the following step. Further details about the knitting process and simulation setup are given in the Materials and Methods section.

In Figure~\ref{fig:knit_patterns} we include simulations of several knit-purl patterns that are ``unbalanced,'' with a larger total density of either knit or purl stitches. This imbalance results in out-of-plane curvatures at scales comparable to (if not larger than) the length scale of the pattern's features. For these cases, we included a set of weak springs in the simulation that connect the mesh's boundary vertices to their initial planar positions, keeping the fabric relatively flat overall and allowing us to emphasize the smaller-scale patterning. Physical samples do sometimes allow for these large-scale curvatures, although for our samples these features were largely mitigated when simply placed on a table or allowed to hang under gravity.

\subsection*{Tuning Parameters}

In our simulations, we chose natural curvatures smaller than the inverse thickness $1/h$ to avoid generating deformations large enough to cause the surface to self-intersect, as our simulation does not penalize self-intersecting surfaces. We also note that self-avoidance forces may be critical to the formation of certain knit-purl buckling patterns in a fully relaxed fabric. Although we used $\kappa_x = \kappa_y$ for simplicity in our model, the exact ratio of curvatures in different directions does not yield significantly different results. Indeed, we expect experimentally that $\kappa_x \neq \kappa_y$ as the knit structure (Figure~\ref{fig:knit_schematic}a) is inherently anisotropic.

As noted previously, the elastic sheet thickness $h$ serves as a tuning parameter relating the fabric's effective bending and tensile moduli, and a suitable value was chosen based on comparisons between initial simulations and experiment. In particular, smaller values of $h$ energetically disfavor regions of nonzero Gaussian curvature, which appear in localized regions of knitted patterns.
For patterns featuring curved ridges (e.g. Figure~\ref{fig:knit_patterns}a,e), too-small values of $h$ therefore result in simulations failing to reproduce the Gaussian curvatures required for such features.

Finally, we note that swatches of pure knit or purl stitches show the Poisson effect, implying $0 < \nu < 0.5$, where $\nu$ is the 3D Poisson ratio of the material \cite{poincloux2018geometry}. In our simulations, we chose $\nu = 0.4$ to approximate the material as having a small amount of compressibility, but variations of $\nu$ between $0$ and $0.5$ have no significant visual effect.

It is essential to note that the buckling patterns generated in Figure~\ref{fig:knit_patterns} are generic and not strongly dependent on the exact parameters $h$, $\kappa_x$, $\kappa_y$, and $\nu$; small changes in these values do not affect the qualitative results for the knit-purl motifs we tested.

\section*{Discussion}

\begin{figure*}[h!]
    \centering
    \includegraphics[width=.65\textwidth]{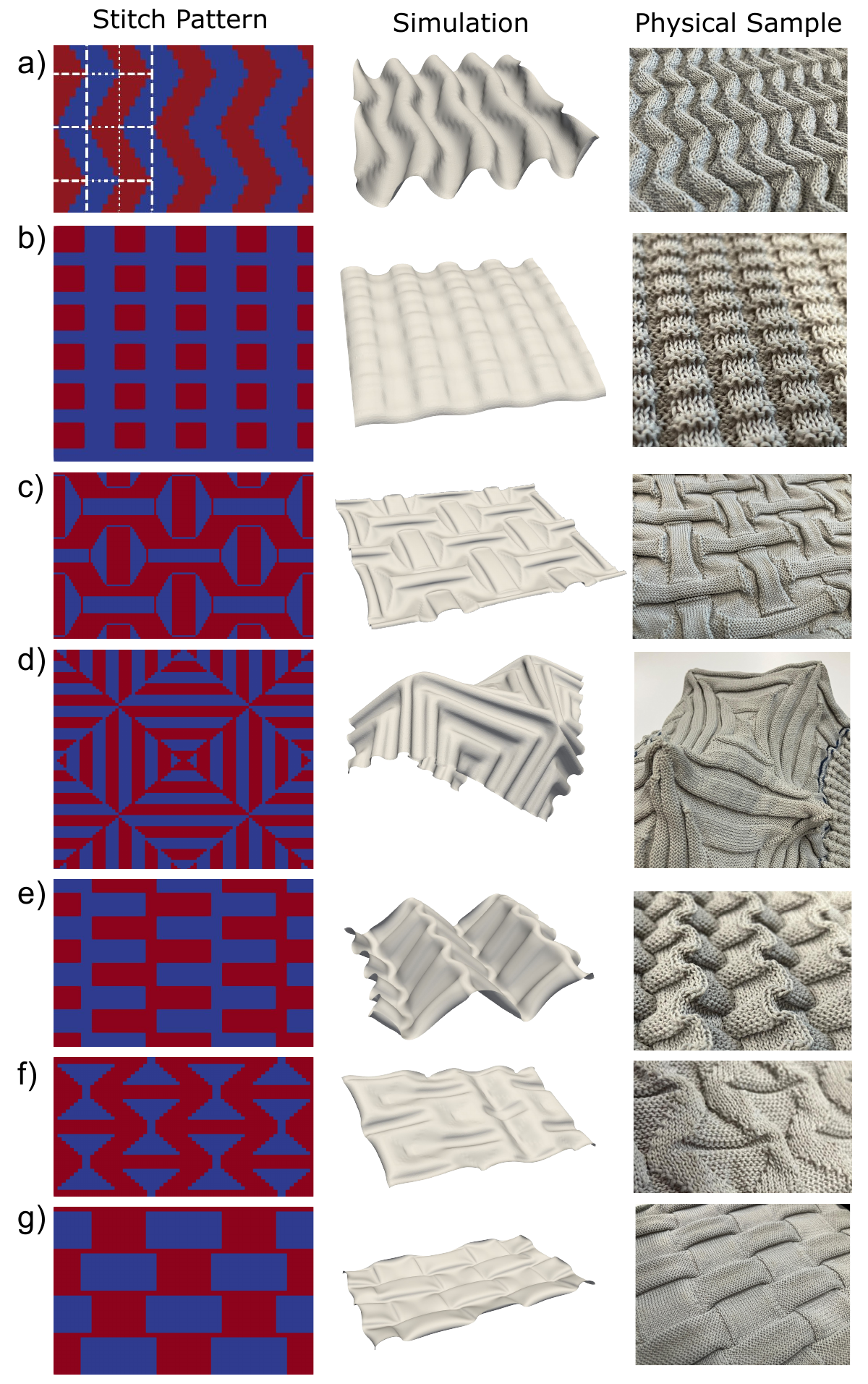}
    \caption{
        (a-g) Comparison of knit-purl patterns (left), simulated meshes (center), and knitted samples (right).
        Red regions represent knit stitches; blue regions represent purl stitches. Note that not all patterns are at the same scale or have the same number of stitches.
        Parameters for all simulations are as follows: each rectangular ``stitch'' in the mesh has dimension $1.4 \times 1.2$ (based on the stitch aspect ratio from knit samples). The elastic thickness parameter $h = 0.5$, and material parameters for the (isotropic) elastic surface approximation are Young's modulus $Y = 1$ and Poisson ratio $\nu = 0.4$. Natural curvatures $\kappa_x = -1$ and $\kappa_y = 1$.
        Pattern in (a) has some fold lines overlaid on the left side to show the crease pattern in the final fabric. Dashed lines represent valley folds; dotted lines represent mountain folds.
        Patterns in (b), (c), (f), and (g) are not symmetric with respect to the density of knit and purl stitches, and with free boundary conditions the meshes will develop curvature at large wavelengths. To mitigate this effect, we use weak springs at boundary vertices with spring constant per unit length $10^{-4}$ between the vertices and their positions on the initial (flat) mesh. Boundary conditions are otherwise all free.
        All experimental samples, besides (d), are slightly stretched in the fabric plane to emphasize their texture near the flat state, and to avoid the amount of self-intersection relevant to the pattern. Simulations did not forbid self-intersection.
    }
    \label{fig:knit_patterns}
\end{figure*}

The problem of designing a self-folding knitted swatch is distinct from the problem of designing folding patterns for traditional origami. In particular, interfaces between regions of knit and purl stitches often do \textit{not} correspond to regions of large fabric curvatures or folds, and the ``creases'' observed in a self-folding knitted fabric may be difficult to guess simply by looking at the stitch pattern itself. Because the natural curvatures $\kappa_x$ and $\kappa_y$ change sign at any knit-purl interface, these interfacial regions typically remain flat. For instance, in the Miura-Ori knitting pattern (Figure~\ref{fig:knit_patterns}a) the observed ``fold lines'' run directly horizontally and vertically instead of following the diagonal interfaces between knit and purl stitches that one might expect from the {\sl origami} structure \cite{luan2020auxetic}.

The final form of the fabric is not in general locally predictable, as its shape is the result of solving a global optimization problem. Therefore, even patterns that are visually similar may yield dramatically different fabric textures. However, we note that all sharp creases and regions of large curvature tend to be cylindrical and aligned either horizontally or vertically, and they follow the axis-aligned principal curvature directions of plain knit or purl swatches. Locally, we can thus view each knit region of the fabric (on the order of a few stitch lengths) as being selected to curl cylindrically in the $x$ direction (negatively), $y$ direction (positively), or remain flat. A purl region of the fabric has similar choices, with opposite curvatures.

\subsection*{Future Improvements}

Although the goal of our analysis is simply to reproduce the initial buckling pattern of fabrics made with knit and purl stitches, several improvements could be made that we have so far ignored in favor of simplicity and parsimony in the initial model.

Figures~\ref{fig:knit_patterns}a,e,f show the tendency for some knit-purl patterns to form sharp creases or corners, with features at length scales comparable to single stitches and the thickness of the fabric. Although we approximate knitted stitches as having no internal structure, such creases emerge frequently in knits. However, simulation of such features likely depends on the geometry of individual stitches, and furthermore requires discretizing the mesh on scales much smaller than the thickness, which becomes computationally costly and fails to capture the obvious variation of the yarn at that scale.

Furthermore, regions of knit and purl stitches meeting at a horizontal or vertical interface are offset slightly relative to each other (in the direction normal to the fabric plane), and these offset features remain even when the fabric is flattened under tension. These offsets are typically at the stitch length scale, but we note that they might have effects on the equilibrium fabric shape and its mechanics. In particular, across these offsets the physical fabric thickness is not well-defined, and the effective bending moduli of these regions may be significantly different from those within a bulk region of knit or purl stitches.

We assumed isotropic elasticity in our model as well as constant thickness and moduli throughout the material. Even swatches made from purely knit stitches are anisotropic in their moduli, and as mentioned other physical moduli need not be assumed homogeneous, particularly across a knit-purl boundary \cite{ruan1996experimental, ko1997effect}. However, our choice of model was partially motivated by the desire to find as simplistic a description as possible of patterning in knit-purl motifs.

Finally, we hope to include an energetic cost to the fabric's self-intersection in future work.  We hope that this work acts as a catalyst for the science of Knitogami\textsuperscript{TM}.

\subsection*{Conclusion}

We have shown that a simple approximation of planar knit-purl patterns as elastic sheets, with natural curvatures defined according to the stitch pattern, can recapitulate the self-folding behavior of several knit-purl motifs. Although the model does not capture quantitative details of the knitted fabric at the stitch length scale, it shows promise in helping to design complex patterns with simple knit structures, as well as motivating a simple framework for understanding knitted fabric geometry.

\section*{Materials and Methods}

\subsection*{Experimental Samples}
All fabric samples used for comparison with simulation were knitted using Unifi REPREVE® Performance nylon yarn (doubled), on a Shima Seiki SSG 122SV two-bed weft knitting machine.

\subsection*{Two Dimensional Surface Simulation}

When we observe the folded fabric we find that the general geometric motifs are independent of the details of the yarn and machine parameters.  With this in mind, we do not expect it to be necessary to provide the specific details of the two-dimensional surface energy.  To connect with the well-known literature on thin-plate theory, we employ the Föppl-von Kármán energy both for ease of interpretation as an effective model, and to exploit existing software developed for simulating thin elastic sheets \cite{vanrees2017growth}.

The Föppl-von Kármán energy, which includes bending and stretching terms for a thin elastic sheet, can be written in terms of the first and second fundamental forms $a$ and $b$ of a flat region $\Omega \in \mathbb{R}^2$ mapped to the three-dimensional midsurface of the sheet in $\mathbb{R}^3$:
\begin{equation}
U = \frac{1}{2} \int_\Omega \left[\frac{h}{4} \left\Vert a_0^{-1} (a - a_0) \right\Vert_e^2 + \frac{h^3}{12} \left\Vert a_0^{-1} (b - b_0) \right\Vert_e^2 \right] \text{d}A
\end{equation}
where the norm $\left\Vert X \right\Vert_e^2 = \frac{Y \nu}{1-\nu^2} \text{tr}^2 X + \frac{Y}{1+\nu} \text{tr}(X^2)$, $a_0$ and $b_0$ represent the initial (zero-energy) fundamental forms of the sheet, and Young's modulus and Poisson ratio are $Y$ and $\nu$ respectively. In this formulation, the Young's modulus becomes a constant factor in the total energy and its value does not affect the minimum energy surface geometry. For flat sheets with no internal stresses, $a_0 = I$ and $b_0 = 0$; for sheets with flat metric but with natural curvatures $\kappa_x$ and $\kappa_y$ in the $x$ and $y$ directions respectively, $b_0 = -\text{diag}\left( \kappa_x, \kappa_y \right)$, a diagonal matrix. For a triangular mesh approximating a continuous surface, the integral is replaced by a sum over triangles. We follow the finite-difference scheme used in \cite{vanrees2017growth} to define $a_t$ and $b_t$ on triangle $t$ as
\begin{equation}
a_t = \begin{pmatrix}
\vec{e}_1 \cdot \vec{e}_1 & \vec{e}_1 \cdot \vec{e}_2 \\
\vec{e}_1 \cdot \vec{e}_2 & \vec{e}_2 \cdot \vec{e}_2
\end{pmatrix},
\quad
b_t = \begin{pmatrix}
2 \vec{e}_1 \cdot \Delta \vec{n}_{20} & -2 \vec{e}_1 \cdot \vec{n}_0 \\
-2 \vec{e}_1 \cdot \vec{n}_0 & 2 \vec{e}_2 \cdot \Delta\vec{n}_{01}
\end{pmatrix}
\end{equation}
with $\vec{e}_i$ as the directed edges of the triangle, $\vec{n}_i$ as the surface normals, and $\Delta \vec{n}_{ij} = \vec{n}_j - \vec{n}_i$. Surface normals are defined on the midpoint of and constrained to be perpendicular to each edge in the mesh, which requires one extra degree of freedom per edge of the mesh and is detailed in \cite{weischedel2012construction}. With these definitions of $a$ and $b$, the total area-weighted Föppl-von Kármán energy of all triangles is minimized over the mesh degrees of freedom (the free vertex positions and edge normals) using the L-BFGS method. Code is available at
\href{https://github.com/jeffersontide/knitogami2024}{https://github.com/jeffersontide/knitogami2024}.

\section{Acknowledgments}

We thank Chelsea E. Amanatides for her extensive work in the initial conception of this project, as well as fabrication of all physical samples. We also thank Michael S. Dimitriyev for helpful comments on a draft of this paper. This work was supported by a Simons Investigator Grant from the Simons Foundation (RDK) and the Kaufman Foundation New Research Initiative award. Effort was also sponsored by the U.S. Government under Other Transaction number HQ00342190016 between Advanced Functional Fabrics of America, Inc. and the Government. The U.S. Government and AFFOA is authorized to reproduce and distribute reprints for Governmental purposes notwithstanding any copyright notation thereon. The views and conclusions contained herein are those of the authors and should not be interpreted as necessarily representing the official policies or endorsements, either expressed or implied, of the U.S. Government or AFFOA. We dedicate this paper to Mark Warner with whom we would have liked to share it.

\end{document}